\documentclass[journal]{IEEEtran}
\usepackage{epsfig}
\usepackage{graphicx}
\usepackage{amsmath}
\usepackage{amssymb}
\usepackage{verbatim}
\usepackage{mathrsfs}
\usepackage{algorithm}
\usepackage[noend]{algpseudocode}
\usepackage{xcolor}
\usepackage{colortbl}
\usepackage{lipsum}
\usepackage{amsmath,amssymb}
\usepackage{amssymb}%
\usepackage{mathrsfs}%
\usepackage{blindtext}

\makeatletter
\def\BState{\State\hskip-\ALG@thistlm}
\makeatother
\usepackage{cite}
\usepackage[T1]{fontenc} 

\IEEEoverridecommandlockouts
\begin{document}
%
\title{Deep Reinforcement Learning based Resource Allocation for V2V Communications}
\author{\IEEEauthorblockN{Hao Ye, Geoffrey Ye Li, and Biing-Hwang Fred Juang 
}\\


\thanks{This work was supported in part by a research
gift from Intel Corporation and the National Science Foundation under Grants
1443894 and 1731017.

H. Ye, G. Y. Li, and B-H. F. Juang are with the School of Electrical and Computer
Engineering, Georgia Institute of Technology, Atlanta, GA 30332 USA (email: yehao@gatech.edu; liye@ece.gatech.edu; juang@ece.gatech.edu)}
}
\maketitle
%
\begin{abstract}

In this paper, we develop a decentralized resource allocation mechanism for vehicle-to-vehicle (V2V) communications based on deep reinforcement learning, which can be applied to both unicast and broadcast scenarios. According to the decentralized resource allocation mechanism,  an autonomous ``agent'', a V2V link or a vehicle, makes its decisions to find the optimal sub-band and power level for transmission without requiring or having to wait for global information. Since the proposed method is decentralized, it incurs only limited transmission overhead. From the simulation results, each agent can effectively learn to satisfy the stringent latency constraints on V2V links while minimizing the interference to vehicle-to-infrastructure (V2I) communications.

\end{abstract}
\begin{IEEEkeywords}
Deep Reinforcement Learning, V2V Communication, Resource Allocation
\end{IEEEkeywords}
\section{Introduction}
\label{sec:intro}
Vehicle-to-vehicle (V2V) communications \cite{Liang_A,Liang_B,VTM} have been developed as a key technology in enhancing the transportation and road safety by supporting cooperation among vehicles in close proximity. Due to the safety imperative,  the quality-of-service (QoS) requirements for V2V communications are very stringent with ultra low latency and high reliability\cite{V2X_service}.
Since proximity based device-to-device (D2D) communications provide direct local message dissemination with substantially decreased latency and energy consumption, the Third Generation Partnership (3GPP) supports V2V services based on D2D communications\cite{3GPP} to satisfy the QoS requirement of V2V applications.

In order to manage the mutual interference between the D2D links and the cellular links, effective resource allocation mechanisms are needed. In \cite{Feng}, a three-step approach has been proposed, where the transmission power is controlled and the spectrum is allocated to maximize the system throughput with constraints on minimum signal-to-interference-plus-noise ratio (SINR) for both the cellular and the D2D links.  In V2V communication networks, new challenges are brought about by high mobility vehicles. As high mobility causes rapidly changing wireless channels, traditional methods on resource management for D2D communications with a full channel state information (CSI) assumption can no longer be applied in the V2V networks.

Resource allocation schemes have been proposed to address the new challenges in D2D-based V2V communications. The majority of them are conducted in a centralized manner, where the resource management for V2V communications is performed in a central controller. In order to make better decisions, each vehicle has to report the local information, including local channel state and interference information, to the central controller. With the collected information from vehicles, the resource management is often formulated as optimization problems, where the constraints on the QoS requirement of V2V links are addressed in the optimization constraints. Nevertheless, the optimization problems are usually NP-hard, and the optimal solutions are often difficult to find. As alternative solutions, the problems are often divided into several steps so that local optimal and sub-optimal solutions can be found for each step. In \cite{Sun}, the reliability and latency requirements of V2V communications have been converted into optimization constraints, which are computable with only large-scale fading information and a heuristic approach has been proposed to solve the optimization problem. In \cite{Le}, a resource allocation scheme has been developed only based on the slowly varying large-scale fading information of the channel and the sum V2I ergodic capacity is optimized with V2V reliability guaranteed.

Since the information of vehicles should be reported to the central controller for solving the resource allocation optimization problem, the transmission overhead is large and grows dramatically with the size of the network, which prevents these methods from scaling to large networks. 
Therefore, in this paper, we focus on decentralized resource allocation approaches, where there are no central controllers collecting the information of the network. In addition, the distributed resource management approaches will be more autonomous and robust, since they can still operate well when the supporting infrastructure is disrupted or become unavailable. 
Recently, some decentralized resource allocation mechanisms for V2V communications have been developed. 
A distributed approach has been proposed in \cite{baseline} for  spectrum allocation for  V2V communications by utilizing the position information of each vehicle. The V2V links are first grouped into clusters according to the positions and load similarity.  The resource blocks (RBs) are then assigned to each cluster and within each cluster, the assignments are improved by iteratively swapping the spectrum assignments of two V2V links. 
In \cite{Low_Com}, a distributed algorithm has been designed to optimize outage probabilities for V2V communications based on bipartite matching.

The above methods have been proposed for unicast communications in vehicular networks. However, in some applications in V2V communications, there is no specific destination for the exchanged messages. In fact, the region of interest for each message includes the surrounding vehicles, which are the targeted destinations. Instead of using a unicast scheme to share the safety information, it is more appropriate to employ a broadcasting scheme. However, blindly broadcasting messages can cause the broadcast storm problem, resulting in package collision. In order to alleviate the problem, broadcast protocols based on statistical or topological information have been investigated in \cite{Storm,Baseline_Pro}. In \cite{Baseline_Pro}, several forwarding node selection algorithms have been proposed based on the distance to the nearest sender, of which the $p$-persistence provides the best performance and is going to be used as a part of our evaluations.

In the previous works, the QoS of V2V links only includes the reliability of SINR and the latency constraints for V2V links has not been considered thoroughly since it is hard to formulate the latency constraints directly into the optimization problems. To address these problems, we use deep reinforcement learning to handle the resource allocation in unicast and the broadcasting vehicular communications. Recently, deep learning has made great stride in  speech recognition \cite{DL_Speech}, image recognition \cite{Imagenet}, and wireless communications \cite{Hao}. With deep learning techniques, reinforcement learning has shown impressive improvement in many applications, such as playing videos games\cite{Atari}, playing Go games\cite{Go}, and job scheduling with multiple resource demands in the computing clusters \cite{DRL_RM}.

Encouraged by our initial results in \cite{Hao_RL}, we exploit deep reinforcement learning to find the mapping between the local observations, including local CSI and interference levels, and the resource allocation and scheduling solution in this paper. In the unicast scenario, each V2V link is considered as an agent and the spectrum and transmission power are selected based on the observations of instantaneous channel conditions and exchanged information shared from the neighbors at each time slot.  Apart from the unicast case, deep reinforcement learning based resource allocation framework can also be extended to the broadcast scenario. In this scenario, each vehicle is considered as an agent and the spectrum and messages are selected according to the learned policy.  In general, the agents will automatically balance between minimizing the interference of the V2V links to the vehicle-to-infrastructure (V2I) links and meeting the requirements for the stringent latency constraints imposed on the V2V link.

The main contribution of this paper is to develop a decentralized resource allocation in V2V communications based on a multi-agent deep reinforcement learning, where the latency constraints on V2V links can be directly addressed. 
The framework can be applied to both unicast and broadcast communication scenarios. Our simulation results demonstrates that deep reinforcement learning based resource allocation can effectively learn to share the channel with V2I and other V2V links and generate the least interference to the V2I links.

The rest of the paper is organized as follows. In Section \ref{Sec:SystemModel}, the system model for unicast communications is introduced. In Section \ref{Sec:ProposedMethod}, the reinforcement learning based resource allocation framework for unicast V2V communications is presented in detail. In Section \ref{Sec:Broadcast}, the framework is extended to the broadcast scenario, where the message receivers are all vehicles within a fixed area. In Section \ref{Sec:Simulation}, the simulation results are presented and the conclusions are drawn in Section \ref{Sec:Conclusion}.

\section{System Model for Unicast Communication} \label{Sec:SystemModel}

\begin{figure}[!t]
 \centering
\includegraphics[width=0.95\linewidth]{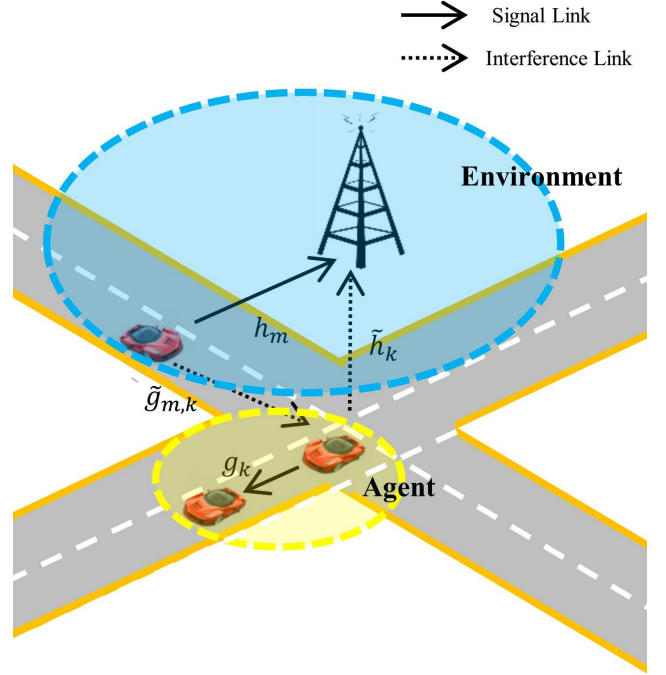}
\caption{An illustrative structure of unicast vehicular communication networks.} \label{fig:System}
\end{figure}

In this section, the system model and resource management problem of unicast communications are presented.  As shown in Fig. \ref{fig:System}, the vehicular networks include  $M$ cellular users (CUEs) denoted by $\mathcal{M} = \{1,2,...,M\}$  and $K$  pairs of V2V users (VUEs) denoted by $\mathcal{K} = \{1,2,...,K\}$. The CUEs demand V2I links to support high capacity communication with the base station (BS) while the VUEs need V2V links to share information for traffic safety management.
To improve the spectrum utilization efficiency, we assume that the orthogonally allocated uplink spectrum for V2I links is shared by the V2V links since the interference at the BS is more controllable and the uplink resources are less intensively used.

The SINR of the $m$th CUE can be expressed as
 \begin{equation}
\gamma^c_m = \frac{P^c_m h_{m}}{\sigma^2 +  \sum_{k\in \mathcal{K}}{\rho_{m,k} P^d_k\tilde{h}_{k}}},
\end{equation}
where $P^c_m$ and $P^d_k$ denotes the transmission powers of the $m$th CUE and the $k$th VUE, respectively, $\sigma^2$ is the noise power, $h_{m}$ is the power gain of the channel corresponding to the $m$th CUE, $\tilde{h}_{k}$ is the interference power gain of the $k$th VUE, and $\rho_{m,k}$ is the spectrum allocation indicator with $\rho_{m,k} = 1$ if the $k$th VUE reuses the spectrum of the $m$th CUE and $\rho_{m,k} = 0$ otherwise.
Hence the capacity of the $m$th CUE is
\begin{equation}
 C^c_m = W \cdot \log(1+\gamma_m),
\end{equation}
 where $W$ is the bandwidth.

Similarly,  the SINR of the $k$th VUE can be expressed as
 \begin{equation}
 \gamma^d_k = \frac{P^d_k\cdot g_{k}}{\sigma^2 + G_c + G_d},
 \end{equation}
 where 
 \begin{equation}
 G_c =  \sum_{m\in \mathcal{M}}{\rho_{m,k} P^c_m\tilde{g}_{m,k}},
 \end{equation}
is the interference power of the V2I link sharing the same RB and
 \begin{equation}
 G_d = \sum_{m\in \mathcal{M}}\sum_{k'\in \mathcal{K} k \neq k'}{\rho_{m,k}\rho_{m,k'}  P^d_{k'}\tilde{g}^d_{k',k}},       
 \end{equation}
is the overall interference power from all V2V links sharing the same RB, $g_{k}$ is the power gain of the $k$th VUE, $\tilde{g}_{k,m}$ is the interference power gain of the $m$th CUE, and $\tilde{g}^d_{k',k}$ is  the interference power gain of the $k'$th VUE. The capacity of the $k$th VUE can be expressed as \begin{equation}
 C^d_k = W \cdot \log(1+\gamma^d_k).
\end{equation}


Due to the essential role of V2V communications in vehicle security protection, there are stringent latency and reliability requirements for V2V links while the data rate is not of great importance. 
The latency and reliability requirements for V2V links are converted into the outage probabilities\cite{Le,Sun} in system design and considerations. With deep reinforcement learning, these constraints are formulated as the reward function directly, in which a negative reward is given when the constraints are violated. In contrast to V2V communications of safety information, the latency requirement on the conventional cellular traffic is less stringent and traditional resource allocation practices based on maximizing the throughput under certain fairness consideration remain appropriate. Therefore, the V2I sum rate will remain a factor in the reward function for maximization in our method, as we can see in Section III

Since the BS has no information on the V2V links,  the resource allocation procedures of the V2I network should be independent of the resource management of the V2V links. Given resource allocation of V2I links, the objective of the proposed resource management scheme is to ensure satisfying the latency constraints for V2V links while minimizing the interference of the V2V links to the V2I links. In the decentralized resource management scenario, the V2V links will select the RB and transmission power based on the local observations.

The first type of observation that relates to resource allocation is the channel and the interference information. The instantaneous channel information of the corresponding V2V link, $\mathbf{G_t} = [G_1, G_2, ..., G_M]$, is a vector of dimension $M$, where each item corresponds to the power gain of the sub-channel of the V2V links. The channel information of the V2I link, $\mathbf{H_t} = [H_1,H_2, ..., H_M]$, characterizes the power gain of each sub-channel from the transmitter to the BS. The received interference signal strength in the previous time slot, $\mathbf{I_{t-1}} = [I_1,I_2, ..., I_M]$,  represents the received interference strength in each sub-channel. Local observations also include information shared by the neighbors, such as the channel indices selected by the neighbors in the previous time slot, $\mathbf{N_{t-1}} = [N_1, N_2, ..., N_M]$, where each item represents the number of times the sub-channel has been used by the neighbors. In addition, information about the condition of transmitted messages should also be involved, such as the remaining load for transmission, $L_t$, i.e., the proportion of bits remaining to transmit, and the time left before violating the latency constraint, $U_t$. This set of local information, including $\mathbf{H_t}$, $\mathbf{I_{t-1}}$, $\mathbf{N_{t-1}}$, and $U_t$, will be used in the broadcast scenario as we will see in Section III.
Nevertheless, the relationship between the observations and the optimal resource allocation solution is often implicit and not easy to establish. Deep reinforcement learning is exploited to approximate the relationship and to accomplish optimization.

\section{Deep Reinforcement Learning for Unicast Resource Allocation} \label{Sec:ProposedMethod}
In this section, the deep reinforcement learning based resource management for unicast V2V communications is introduced. The formulations of key parts in the reinforcement learning are shown and the deep Q-network based proposed solution is presented in detail.


\subsection{Reinforcement Learning }
\begin{figure}[!t]
\centering
\includegraphics[width = 1.05 \linewidth]{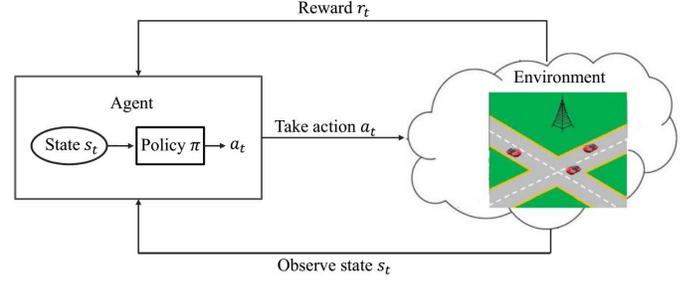}
\caption{Deep reinforcement learning for V2V communications} \label{fig:DRL}
\end{figure}

As shown in Fig. \ref{fig:DRL}, the framework of reinforcement learning consists of agents and environment interacting with each other. In this scenario, each V2V link is considered as an agent and everything beyond the particular V2V link is regarded as the environment, which presents a collective rather than atomized condition related to the resource allocation.
Since the behavior of other V2V links cannot be controlled in the decentralized setting, the action of each agent (individual V2V links) is thus based on the collective manifested environmental conditions such as spectrum, transmission power, etc.

As in Fig. \ref{fig:DRL}, at each time $t$, the V2V link, as the agent, observes a state, $\mathbf{s_t}$, from the state space, $\mathcal{S}$, and accordingly takes an action, $a_t$, from the action space, $\mathcal{A}$,  selecting sub-band and transmission power based on the policy, $\pi$. The decision policy, $\pi$, can be determined by the state-action function, also called Q-function, $Q(\mathbf{s_t}, a_t)$, which can be approximated by deep learning. 
Based on the actions taken by the agents, the environment transits to a new state, $\mathbf{s_{t+1}}$,  and each agent receives a reward, $r_t$, from the environment. In our case, the reward is determined by the capacities of the V2I and V2V links and the latency constraints of the corresponding V2V link. 

As we have discussed in Section II, the state of the environment observable to by each V2V link consists of several parts: the instantaneous channel information of the corresponding V2V link, $\mathbf{g_t}$, the previous interference power to the link, $\mathbf{I_{t-1}}$, the channel information of the V2I link, e.g., from the V2V transmitter to the BS, $\mathbf{h_t}$, the selected of sub-channel of neighbors in the previous time slot, $\mathbf{N_{t-1}}$, the remaining load of the VUE to transmit, $L_t$ , and the remaining time to meet the latency constraints $U_t$. In summary, the state can be expressed as $\mathbf{s_t} = [ \mathbf{I_{t-1}}, \mathbf{H_t}, \mathbf{N_{t-1}}, U_t, \mathbf{G_t},L_t]$.  

At each time, the agent takes an action $a_t \in \mathcal{A}$, which consists of selecting a sub-channel and a power level for transmission, based on the current state, $\mathbf{s_t} \in \mathcal{S}$, by following a policy $\pi$. The transmission power is discretized into three levels, thus the dimension of the action space is $3 \times N_{RB}$ when there are $N_{RB}$ resource blocks in all.

The objective of V2V resource allocation is as follows. An agent (i.e. a V2V link) selects the frequency band and transmission power level that incur only small interference to all V2I links as well as other V2V links while preserving enough resources to meet the requirement of the latency constraints.
Therefore, the reward function consists of three parts, namely, the capacity of the V2I links, the capacity of the V2V links, and the latency condition. The sum capacities of the V2I and the V2V links are used to measure the interference to the V2I and other V2V links, respectively. The latency condition is represented as a penalty. In particular, the reward function is expressed as,
\begin{equation}
r_t = \lambda_c\sum_{m\in \mathcal{M} }{C^c_m} + \lambda_d\sum_{k \in \mathcal{K}}{C^d_k} - \lambda_p(T_0 - U_t),
\end{equation}
where $T_0$ is the maximum tolerable latency and $\lambda_c$, $\lambda_d$, and $\lambda_p$ are weights of the three parts. The quantity $(T_0 - U_t)$ is the time used for transmission; the penalty increases as the time used for transmission grows. In order to obtain a good performance in the long-term, both the immediate rewards and the future rewards should be taken into consideration. Therefore, the main objective of reinforcement learning is to find a policy to maximize the expected cumulative discounted rewards,
\begin{equation}
G_t = \mathbb{E}[\sum_{n=0}^{\infty} \beta^n r_{t+n}],
\end{equation}
where $\beta \in [0,1]$ is the discount factor.

The state transition and reward are stochastic and modelled as a Markov decision process (MDP), where the state transition probabilities and rewards depend only on the state of the environment and the action taken by the agent. The transition from $\mathbf{s_t}$ to $\mathbf{s_{t+1}}$ with reward $r_t$ when action $a_t$ is taken can be characterized by the conditional transition probability, $p(\mathbf{s_{t+1}},r_t|\mathbf{s_t},a_t)$. 
It should be noted that the agent can control its own actions and has no prior knowledge on the transition probability matrix $\mathbf{P} = \{p(\mathbf{s_{t+1}},r_t|\mathbf{s_t},a_t)\}$, which is only determined by the environment.
In our problem, the transition on the channels, the interference, and the remaining messages to transmit are generated by the simulator of the wireless environment.

\subsection{Deep Q-Learning}
The agent takes actions based on a policy, $\pi$, which is a mapping from the state space, $\mathcal{S}$, to the action space, $\mathcal{A}$,  expressed as $\pi: \mathbf{s_t} \in \mathcal{S} \rightarrow a_t \in \mathcal{A}$.
As indicated before, the action space spans over two dimensions, the power level and the spectrum subband, and a action, $a_t \in \mathcal{A}$, corresponds to a selection of the power level and the spectrum for V2V links.

The Q-learning algorithms can be used to get an optimal policy to maximize the long-term expected accumulated discounted rewards, $G_t$ \cite{Deep}. 
The Q-value for a given state-action pair $(\mathbf{s_t},a_t)$, $Q(\mathbf{s_t}, a_t)$, of policy $\pi$ is defined as the expected accumulated discounted rewards when taking an action $a_t \in \mathcal{A}$ and following policy $\pi$ thereafter.
Hence the Q-value can be used to measure the \emph{quality} of certain action in a given state.
Once Q-values, $Q(\mathbf{s_t}, a_t)$, are given, an improved policy can be easily constructed by taking the action, given by 
\begin{equation}
a_t = \arg\max_{a \in \mathcal{A}}Q(\mathbf{s_t},a).
\end{equation}
That is, the action to be taken is the one that maximizes the Q-value.

The optimal policy with Q-values $Q^*$ can be found without any knowledge of the system dynamics based on the following update equation,
\begin{equation}
\begin{aligned}
Q_{new}(\mathbf{s_t}, a_t)  = & Q_{old}(\mathbf{s_t}, a_t) + \alpha [r_{t+1} +\\
                     &\beta \max_{\mathbf{s}\in \mathcal{S}} Q_{old}(\mathbf{s}, a_t) - Q_{old}(\mathbf{s
_t}, a_t)],
\end{aligned}
\end{equation}
It has been shown in \cite{ReinforcementLearning} that in the MDP, the Q-values will converge with probability $1$ to the optimal $Q^*$ if each action in the action space is executed under each state for an infinite number of times on an infinite run and the learning rate $\alpha$ decays appropriately. The optimal policy, $\pi^*$, can be found once the optimal Q-value,  $Q^*$ , is determined.

In the resource allocation scenario, once the optimal policy is found through training, it can be employed to select spectrum band and transmission power level for V2V links to maximize overall capacity and ensure the latency constraints of V2V links.

 \begin{figure}[!t]
\centering
\includegraphics[width = 1.05 \linewidth]{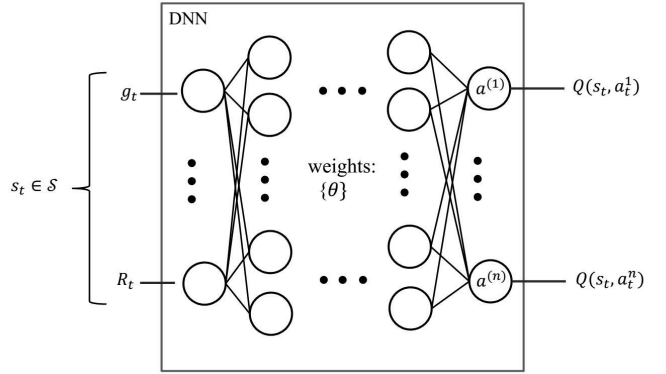} 
\caption{Structure of Deep Q-networks.}\label{fig:DeepQNet}
\end{figure}
The classic Q-learning method can be used to find the optimal policy when the state-action space is small, where a look-up table can be maintained for updaing the Q-value of each item in the state-action space.  
However, the classic Q-learning can not be applied if the state-action space becomes huge, just as in the resource management for the V2V communications.  The reason is that a large number of states will be visited infrequently  and corresponding Q-value will be updated rarely, leading to a much longer time for the Q-function to converge. To remedy this problem, deep Q-network improves the Q-learning by combining the deep neural networks (DNN) with Q-learning. As shown in Fig. \ref{fig:DeepQNet},  the Q-function is approximated by a DNN with weights \{$\boldsymbol{\theta}$\} as a Q-network \cite{Deep}.
Once \{$\boldsymbol{\theta}$\} is determined, Q-values, $Q(\mathbf{s_t}, a_t)$, will be the outputs of the DNN in Fig. \ref{fig:DeepQNet}.
DNN can address sophisticate mappings between the channel information and the desired output based on a large amount of training data, which will be used to determine Q-values.

The Q-network updates its weights, $\boldsymbol{\theta}$, at each iteration to minimize the following loss function derived from the same Q-network with old weights on a data set $D$,
\begin{equation}
Loss(\boldsymbol{\theta}) = \sum_{(\mathbf{s_t},a_t) \in D}(y - Q(\mathbf{s_t},a_t,\boldsymbol{\theta}))^2,
\end{equation}
where 
\begin{equation}
y = r_t+\max_{a \in \mathcal{A}} Q_{old}(\mathbf{s_t},a,\boldsymbol{\theta}),
\end{equation}
where $r_t$ is the corresponding reward.

\subsection{Training and Testing Algorithms}
Like most machine learning algorithms, there are two stages in our proposed method, i.e., the training and the test stage. 
Both the training and test data are generated from the interactions of an environment simulator and the agents. Each training sample used for optimizing the deep Q-network includes $\mathbf{s_t}$, $\mathbf{s_{t+1}}$, $a_t$, and $r_t$. 
The environment simulator includes VUEs and CUEs and their channels, where positions of the vehicles are generated randomly and the CSI of V2V and V2I links is generated according to the positions of the vehicles. 
With the selected spectrum and power of V2V links, the simulator can provide  $\mathbf{s_{t+1}}$ and $r_t$ to the agents. 
In the training stage, the deep Q-learning with experience replay is employed \cite{Deep}, where the training data is generated and stored in a storage named \emph{memory}. 
As shown in Algorithm 1, in each iteration, a mini-batch data is sampled from the \emph{memory} and is utilized to renew the weights of the deep Q-network. 
In this way, the temporal correlation of generated data can be suppressed.  
The policy used in each V2V link for selecting spectrum and power is random at the beginning and gradually improved with the updated Q-networks.
As shown in Algorithm 2, in the test stage, the actions in V2V links are chosen with the maximum Q-value given by the trained Q-networks, based on which the evaluation is obtained.

As the action is selected independently based on the local information, the agent will have no knowledge of actions selected by other V2V links if the actions are updated simultaneously. As a consequence, the states observed by each V2V link cannot fully characterize the environment.
In order to mitigate this issue, the agents are set to update their actions asynchronously, where only one or a small subset of V2V links will update their actions at each time slot.
In this way, the environmental changes caused by actions from other agents will be observable.

\begin{algorithm}
\caption{Training Stage Procedure of Unicast}\label{euclid}
\begin{algorithmic}[1]
\Procedure{Training}{}\\
\textbf{Input}: Q-network structure, environment simulator.\\
\textbf{Output}: Q-network\\

\textbf{Start:}

Random initialize the policy $\pi$ 

Initialize the model

Start environment simulator, generate vehicles, V2V links, V2I links.\\
\textbf{Loop}:

Iteratively select the V2V link in the system.

For each V2V links, select the spectrum and power for transmission based on policy $\pi$.

Environment simulator generates states and rewards based on the action of agents.

Collect and save the data item \{state, reward, action, post-state\} into memory.

Sample a mini-batch of data from the memory.

Train the deep Q-network using the mini-batch data.

Update the policy $\pi$: chose the action with maximum Q-value.\\

\textbf{End Loop}\\

\textbf{Return}: Return the deep Q-network

\EndProcedure
\end{algorithmic}
\end{algorithm}

\begin{algorithm}
\caption{Test Stage Procedure of Unicast}
\begin{algorithmic}[1]
\Procedure{Testing}{}\\
\textbf{Input}: Q-network, environment simulator.\\
\textbf{Output}: Evaluation results\\
\textbf{Start:}
Load the Q-network model

Start environment simulator, generate vehicles, V2V links, V2I links.\\ 

\textbf{Loop}:

Iteratively select a V2V link in the system.

For each V2V link, select the action by choosing the action with the largest Q-value.

Update the environment simulator based on the actions selected.

Update the evaluation results, i.e., the average of V2I capacity and the probability of successful VUEs.  
\\

\textbf{End Loop}\\
\textbf{Return}: Evaluation results

\EndProcedure
\end{algorithmic}
\end{algorithm}

\section{Resource Allocation for Broadcast System} \label{Sec:Broadcast}

\begin{figure}[!t]
\centering
\includegraphics[width=0.95\linewidth]{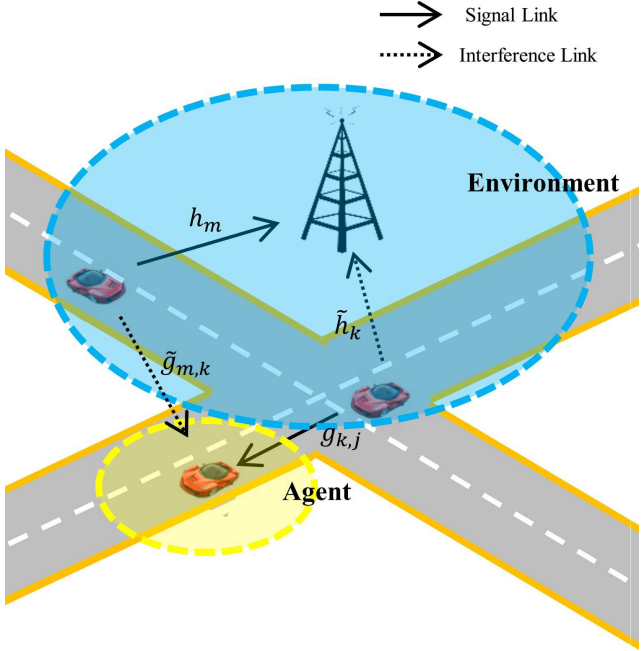}
\caption{An illustrative structure of broadcast vehicular communication networks.} \label{fig:SystemModel_B}
\end{figure}

In this section, the resource allocation scheme based on deep reinforcement learning is extended to the broadcast V2V communication scenario. We first introduce the broadcast system model. Then the key elements in reinforcement learning framework are formulated for the broadcast system and algorithms to train the deep Q-networks are shown.

\subsection{Broadcast System} 

Fig. \ref{fig:SystemModel_B} shows a broadcast V2V communication system, where the vehicle network consists $\mathcal{M_B} = \{1,2,...,M_B\}$ CUEs demanding V2I links. At the same time, $\mathcal{K_B} = \{1,2,...,K_B\}$ VUEs are broadcasting the messages, where each message has one transmitter and a group of receivers in the surrounding area.
Similar to the unicast case, the uplink spectrum for the V2I links is reused by the V2V links as uplink resources are less intensively used and interference at the BS is more controllable.

In order to improve the reliability of broadcast, each vehicle will rebroadcast the messages that have been received. However, as mentioned in Section I, the broadcast storm problem occurs when the density of the vehicles is large and there are excessive redundant rebroadcast messages exist in the networks. To address this problem, vehicles need to select a proper subset of the received messages to rebroadcast so that more receivers can get the messages within the latency constraint while bringing little redundant broadcast to the networks.

The interference to the V2I links comes from the background noise and the signals from the VUEs that share the same sub-band. Thus the capacity of V2I link can be expressed as 
\begin{equation}
\gamma^c_m = \frac{P^c h_{m}}{\sigma^2 +  \sum_{k\in \mathcal{K_B}}{\rho_{m,k} P^d\tilde{h}_{k}}},
\end{equation}
where $P^c$ and $P^d$ are the transmission powers of the CUE and the VUE, respectively, $\sigma^2$ is the noise power, $h_{m}$ is the power gain of the channel corresponding to the $m$th CUE, $\tilde{h}_{k}$ is the interference power gain of the $k$th VUE, and $\rho_{m,k}$ is the spectrum allocation indicator with $\rho_{m,k} = 1$ if the $k$th VUE reuses the spectrum of the $m$th CUE and $\rho_{m,k} = 0$ otherwise.
Hence the capacity of the $m$th CUE can be expressed as
\begin{equation}
 C^c_m = W \cdot \log(1+\gamma_m),
\end{equation}
 where $W$ is the bandwidth.

Similarly, for the $j$th receiver of the $k$th VUE, the SINR is
\begin{equation}
\gamma^d_{k,j} = \frac{P^c g_{k,j}}{\sigma^2 + G_c + G_d},
\end{equation}
 with
 \begin{equation}
 G_c =  \sum_{m\in \mathcal{M_B}}{\rho_{m,k} P^c\tilde{g}_{m,k}},
 \end{equation}
 and
 \begin{equation}
 G_d = \sum_{m\in \mathcal{M_B}}\sum_{k'\in \mathcal{K_B} k \neq k'}{\rho_{m,k}\rho_{m,k'}  P^d\tilde{g}^d_{k',k,j}},   
 \end{equation}
 where $ g_{k,j}$ is the power gain of the $j$th receiver of the $k$th VUE, $\tilde{g}_{k,m}$ is the interference power gain of the $m$th CUE, and $\tilde{g}^d_{k',k,j}$ is  the interference power gain of the $k'$th VUE. The capacity for the $j$th receiver of the $k$th VUE can be expressed as \begin{equation}
 C^d_{k,j} = W \cdot \log(1+\gamma^d_{k,j}).
\end{equation}

In the decentralized settings, each vehicle will determine which messages to broadcast and select which sub-channel to make better use of the spectrum. These decisions are based on some local observations and should be independent of the V2I links. Therefore, after resource allocation procedure of the V2I communications, the main goal of the proposed autonomous scheme is to ensure that the latency constraints for the VUEs can be met while the interference of the VUEs to the CUEs should be minimized. The spectrum selection and message scheduling should be managed according to the local observations.

In addition to the local information used in the unicast case, some information is useful and unique in the broadcast scenario, including the number of times that the message has been received by the vehicle, $O_t$, and the minimum distance to the vehicles that have broadcast the message, $D_t$. $O_t$ can be obtained by maintaining a counter for each message received by vehicle, where the counter will increase one when the message has been received again. If the message has been received from different vehicles, $D_t$ is the minimum distance to the message senders.  In general, the probability of rebroadcasting a message decreases if the message has been heard many times by the vehicle or the vehicle is near to another vehicle that had broadcast the message before.

\subsection{Reinforcement Learning for Broadcast}

Under the broadcast scenario, each vehicle is considered as an agent in our system.  At each time $t$, the vehicle observes a state, $\mathbf{s_t}$, and accordingly takes an action, $a_t$, selecting sub-band and messages based on the policy, $\pi$.  Following the action, the state of the environment transfers to a new state $\mathbf{s_{t+1}}$ and the agent receives a reward, $r_t$, determined by the capacities of the V2I and V2V links and the latency constraints of the corresponding V2V message. 

Similar to the unicast case, the state observed by each vehicle for each received message consists of several parts: the instantaneous channel interference power to the link, $\mathbf{I_{t-1}}$, the channel information of the V2I link, e.g., from the V2V transmitter to the BS, $\mathbf{H_t}$, the selection of sub-channel of neighbors in the previous time slot, $\mathbf{N_{t-1}}$, and the remaining time to meet the latency constraints, $U_t$. Different from the unicast case, we have to include the number of times that the message have been received by the vehicle, $O_t$ and the minimum distance to the vehicles that have broadcast the message, $D_t$, in the state representation,  In summary, the state can be expressed as $\mathbf{s_t} = [\mathbf{I_{t-1}}, \mathbf{H_t}, \mathbf{N_{t-1}},U_t, O_t, D_t]$.

At each time $t$, the agent takes an action at $a_t \in \mathcal{A}$, which includes determining the massages for broadcasting and the sub-channel for transmission. For each message, the dimension of action space is the $N_{RB} + 1$, where $N_{RB}$ is the number of resource blocks. If the agent takes an action from the first $N_{RB}$ actions, the message will be broadcast immediately in the corresponding sub-channel. Otherwise, the message will not be broadcast at this time if the agent takes the last action.

Similar to the unicast scenario, the objective of selecting channel and message for transmission is to minimize the interference to the V2I links with the latency constraints for VUEs guaranteed. In order to reach this objective, the frequency band and messages selected by each vehicle should have small interference to all V2I links as well as other VUEs. It also needs to meet the requirement of latency constraints. Therefore, similar to the unicast scenario, the reward function consists of three parts, the capacity of V2I links, the capacity of V2V links, and the latency condition. To suppress the redundant rebroadcasting, only the capacities of receivers that have not received the message are taken into consideration. Therefore, no capacity of V2V links is added, if the message to rebroadcast has already been received by all targeted receivers.  The latency condition is represented as a penalty if the message has not been received by all the targeted receivers, which increases linearly as the remaining time $U_t$ decreases. Therefore, the reward function can be expressed as,
\begin{equation}
r_t = \lambda_c\sum_{m\in \mathcal{M} }{C^c_m} + \lambda_d\sum_{k \in \mathcal{K}, j \not \in E\{k\}}{C^d_{k,j}} - \lambda_p(T_0 - U_t),
\end{equation}
where $E\{k\}$ represents the targeted receivers that have received the transmitted message.


In order to get the optimal policy, deep Q-network is trained to approximate the Q-function. The training and test algorithm for broadcasting are very similar to the unicast algorithms, as shown in Algorithm 3 and Algorithm 4.

\begin{algorithm}
\caption{Training Stage Procedure of Broadcast}\label{euclid}
\begin{algorithmic}[1]
\Procedure{Training}{}\\
\textbf{Input}: Q-network structure, environment simulator.\\
\textbf{Output}: Q-network\\

\textbf{Start:}

Random initialize the policy $\pi$ 

Initialize the model

Start environment simulator, generate vehicles, VUE, CUE.\\
\textbf{Loop}:

Iteratively select a vehicle in the system.

For each vehicle, select the messages and spectrum for transmission based on policy $\pi$ 

Environment simulator generates states and rewards based on the action of agents.

Collect and save the data item \{state, reward, action, post-state\} into memory.

Sample a mini-batch of data from the memory.

Train the deep Q-network using the mini-batch data.

Update the policy $\pi$: chose the action with maximum Q-value.\\

\textbf{End Loop}\\

\textbf{Return}: Return the deep Q-network

\EndProcedure
\end{algorithmic}
\end{algorithm}

\begin{algorithm}
\caption{Test Stage Procedure of Broadcast}
\begin{algorithmic}[1]
\Procedure{Testing}{}\\
\textbf{Input}: Q-network, environment simulator.\\
\textbf{Output}: Evaluation results\\
\textbf{Start:}
Load the Q-network model

Start environment simulator, generate vehicles, V2V links, V2I links.\\ 

\textbf{Loop}:

Iteratively select a vehicle in the system.

For each vehicle, select the messages and spectrum by choosing the action with the largest Q-value.

Update the environment simulator based on the actions selected.

Update the evaluation results, i.e., the average of V2I capacity and the probability of successful VUEs.  
\\

\textbf{End Loop}\\
\textbf{Return}: Evaluation results

\EndProcedure
\end{algorithmic}
\end{algorithm}

\section{Simulations}\label{Sec:Simulation}

In this section, we present simulation results to demonstrate the performance of the proposed method for unicast and broadcast vehicular communications.

\subsection{Unicast}
We consider a single cell system with the carrier frequency of $2$ GHz. 
We follow the simulation setup for the Manhattan case detailed in 3GPP TR $36.885$ \cite{3GPP}, where there are 9 blocks in all and with both line-of-sight (LOS) and non-line-of-sight (NLOS) channels. 
The vehicles are dropped in the lane randomly according to the spatial Poisson process and each plans to communicate with the three nearby vehicles. Hence, the number of V2V links, $K$, is three times of the number of vehicles. 
Our deep Q-network is a five-layer fully connected neural network with three hidden layers.
The numbers of neurons in the three hidden layers are 500, 250, and 120, respectively.
The activation function of Relu is used, defined as
\begin{equation}
f_r(x) = \max(0,x).
\end{equation}
The learning rate is $0.01$ at the beginning and decreases exponentially.
We also utilize $\epsilon$-greedy policy to balance the exploration and exploitation \cite{Deep} and adaptive moment estimation method (Adam) for training \cite{ADMM}. The detail parameters can be found in Table 1.

\begin{table}
\centering
\caption{Simulation Parameters}
\begin{tabular}{|c|c|}
\hline
\rowcolor{gray}
Parameter & Value\\    \hline
Carrier frequency & 2 GHz \\ \hline
BS antenna height & 25m \\ \hline
BS antenna gain & 8dBi\\ \hline
BS receiver noise figure & 5dB \\ \hline
Vehicle antenna height & 1.5m \\ \hline
Vehicle antenna gain & 3dBi\\ \hline 
Vehicle receiver noise figure  & 9dB \\ \hline
Vehicle speed & 36 km/h\\ \hline
Number of lanes & 3 in each direction (12 in total)\\ \hline
Latency constraints for V2V links $T_0$ & 100 ms\\ \hline
V2V transmit power level list in unicast & [23, 10, 5] dBm\\ \hline
Noise power $\sigma^2$ & -114 dBm\\ \hline
[$\lambda_c$,$\lambda_d$,$\lambda_p$] & [0.1, 0.9, 1] \\ \hline
V2V transmit power in broadcast  & 23 dBm\\ \hline
SINR threshold in broadcast  &  1 dB\\ \hline

\end{tabular}
\label{tab:default}

\end{table}

The proposed method is compared with other two methods. 
The first is a random resource allocation method. 
At each time, the agent randomly chooses a sub-band for transmission.
The other method is developed in \cite{baseline}, where vehicles are first grouped by the similarities and then the sub-bands are allocated and adjusted iteratively to the V2V links in each group.

\begin{figure}[!t]
\centering
\includegraphics[width=1\linewidth]{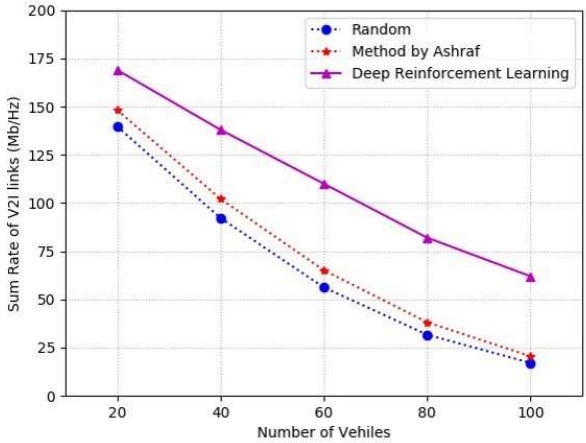}
\caption{Mean rate versus the number of vehicles. }\label{fig:V2I}
\end{figure}

\subsubsection{V2I Capacity}
Fig. \ref{fig:V2I} shows the summation of V2I rate versus the number of vehicles.
From the figure, the proposed method has a much better performance to mitigate the interference of V2V links to the V2I communications. 
Since the method in \cite{baseline} maximizes the SINR in the V2V links, rather than optimizing the V2I links directly, it has only a slightly better performance than the random method, much worse than the proposed method.

\begin{figure}[!t]
\centering
\includegraphics[width=1\linewidth]{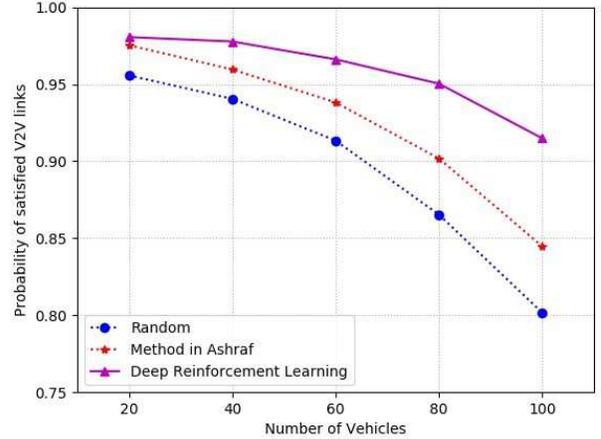}
\caption{Probability of satisfied V2V links versus the number vehicles.}\label{fig:V2V}
\end{figure}

\subsubsection{V2V Latency}
Fig. \ref{fig:V2V} shows the probability that the V2V links satisfy the latency constraint versus the number of vehicles. From the figure, the proposed method has a much larger probability for the V2V links to satisfy the latency constraint since it can dynamically adjust the power and sub-band for transmission so that the links likely violating the latency constraint have more resources.

\subsection{Broadcast}

The simulation environment is same as the one used in unicast, except that each vehicle communicates with all other vehicles within certain distance. The message of the $k$th VUE is considered to be successfully received by the $j$th receiver if the SINR of the receiver, $\gamma^d_{k,j}$, is above the SINR threshold.  If all the targeted receivers of the message have successfully received the message, this V2V transmission is considered as successful.

The deep reinforcement learning based method jointly optimizes the scheduling and channel selection while historical works usually treat the two problems separately.  In the channel selection part, the proposed method is compared with the channel selection method based on \cite{baseline}. In the scheduling part, the proposed method is compared with the $p$-persistence protocol for broadcasting, where the probability of broadcasting is determined by the distance to the nearest sender \cite{Baseline_Pro}.

\subsubsection{V2V Latency}

\begin{figure}[!t]
\centering
\includegraphics[width=1\linewidth]{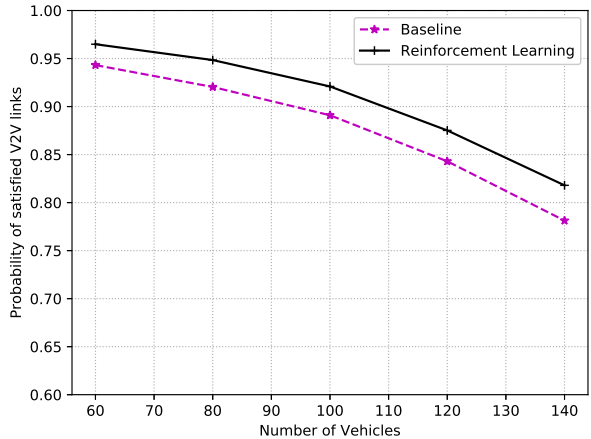}
\caption{Probability of satisfied VUEs versus the number vehicles.}\label{fig:V2V_B}
\end{figure}

Fig. \ref{fig:V2V_B} shows the probability that the VUEs satisfy the latency constraint versus the number of vehicles. From the figure, the proposed method has a larger probability for VUEs to satisfy the latency constraint since it can effectively select the messages and sub-band for transmission.

\subsubsection{V2I Capacity}
\begin{figure}[!t]
\centering
\includegraphics[width=1\linewidth]{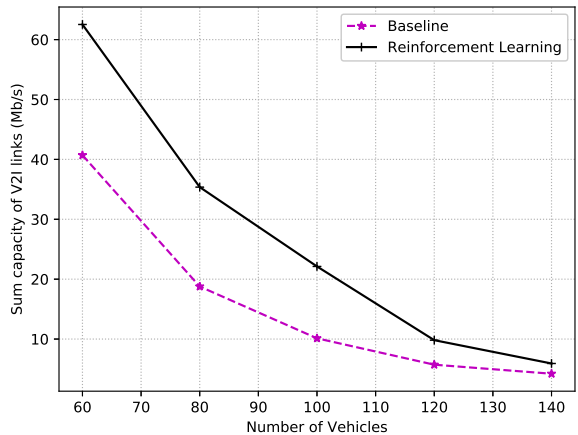}
\caption{Sum capacity versus the number of vehicles. }\label{fig:V2I_B}
\end{figure}

Fig. \ref{fig:V2I_B} demonstrates the summation of V2I rate versus the number of vehicles.
From the figure, the proposed method has a better performance to mitigate the interference of the V2V links to the V2I communications.

\section{Conclusion} \label{Sec:Conclusion}

In this paper, a decentralized resource allocation mechanism has been proposed for the V2V communications based on deep reinforcement learning. It can be applied to both unicast and broadcast scenarios.
Since the proposed methods are decentralized, the global information is not required for each agent to make its decisions, the transmission overhead is small. From the simulation results, each agent can learn how to satisfy the V2V constraints while minimizing the interference to V2I communications.

\section{Acknowledgment}
The authors would like to thank Dr. May Wu, Dr. Satish C. Jha, Dr. Kathiravetpillai Sivanesan, Dr. Lu Lu, and Dr. JoonBeom Kim from Intel Corporation for their insightful comments, which have substantially improved the quality of this paper.

\bibliographystyle{IEEEbib}
\bibliography{strings,refs}

\end{document}